\documentclass[prd,onecolumn,english]{revtex4}
\usepackage{babel}

\usepackage{amsmath}
\usepackage{amssymb}
\usepackage{amsthm} 

\usepackage{array}

\usepackage{slashed}

\usepackage[dvips]{graphicx} 
\usepackage{subfigure}

\usepackage{varioref}   
\usepackage{xr}         

\usepackage{pstricks}           

\allowdisplaybreaks[1]

\newcommand\nn{\nonumber}
\newcommand\ba{\begin{eqnarray}}
\newcommand\ea{\end{eqnarray}}

\newcommand{\Tr}{\mathop{\mathrm{Tr}}\nolimits}

\begin{document}

\title{
\vskip-3cm{\baselineskip14pt
\centerline{\normalsize\rm DESY 12--102\hfill ISSN 0418-9833}
\centerline{\normalsize\rm June 2012\hfill}}
\vskip1.5cm
Production of one or two vector mesons in peripheral high-energy collisions of
heavy ions}

\author{A.~I.~Ahmadov}
\email{ahmadov@theor.jinr.ru}
\affiliation{Bogoliubov Laboratory of Theoretical Physics,
Joint Institute for Nuclear Research, Dubna, 141980  Russia}
\affiliation{Institute of Physics, Azerbaijan National Academy of Science, Baku, Azerbaijan}

\author{B.~A.~Kniehl}
\email{kniehl@desy.de}
\affiliation{{II.} Institut f\"ur Theoretische Physik,
Universit\"at Hamburg, Luruper Chaussee 149, 22761 Hamburg, Germany}

\author{E.~A.~Kuraev}
\email{kuraev@theor.jinr.ru}
\affiliation{Bogoliubov Laboratory of Theoretical Physics,
Joint Institute for Nuclear Research, Dubna,  141980 Russia}

\author{E.~S.~Scherbakova}
\email{scherbak@mail.desy.de}
\affiliation{{II.} Institut f\"ur Theoretische Physik,
Universit\"at Hamburg, Luruper Chaussee 149, 22761 Hamburg, Germany}


\date{\today}

\begin{abstract}
We study the production of spin-one mesons in high-energy heavy-ion collisions
with peripheral kinematics in the framework of QED.
The cross sections of the production of a single vector meson and of two
different ones are presented.
The explicit dependence on the virtuality of the intermediate vector meson is
obtained within a quark model.
The effect of reggeization of the intermediate vector meson state in the case
of the production of two vector mesons is taken into account.

\medskip


\end{abstract}

\maketitle

\section{Introduction}
\label{SectionIntroduction}
The CERN Large Hadron Collider (LHC) provides the opportunity to study
experimentally the production of scalar, pseudo-scalar, and vector mesons in
peripheral collision of heavy ions.
Peripheral kinematics implies the detection of particles produced in directions
close to the axis of the colliding beams \cite{ECHAJA}.
The main feature of these processes is the non-decreasing total and
differential cross sections.
The invariant mass of the created particles is assumed to be small in the
fragmentation and central regions in comparison with the total energy in the
center of mass of the colliding beams $\sqrt{s}=2E$.
The application of the known theoretical approaches, such as the
Nambu-Iona-Lasinio model as well as chiral perturbation theory, seems to be
legitimate.
We shall consider, in pure quantum electrodynamics (QED), the processes of the
production of a single vector meson and of two vector mesons separated by a
rapidity gap.
It is known that the main contributions to the amplitudes of peripheral
processes arise from the interaction mechanism of ions, mediated by the
exchange of spin-one particles, such as virtual photons, vector mesons, and
gluons.
For sufficiently large electric charges of the ions, virtual-photon exchanges
will eventually play the dominant role.
Actually, the effect of the replacement $\alpha\to Z\alpha$ for the case of
charged heavy ions, e.g.\ for Pb-Pb collisions, exceeds the corresponding
QCD contribution for typical regions of momentum transfer, where
$\alpha_s\sim 0.1-0.2$, which seems to be essential in the experimental set-up.
We shall consider processes of the creation of one and two vector mesons, such
as $\omega$, $J/\psi$, $\rho$, or ortho-positronium:
\ba
Y_1(Z_1,P_1)+Y_2(Z_2,P_2)&\to& V(e,r)+Y_1(Z_1,P_1')+Y_2(Z_2,P_2'), \nn \\
Y_1(Z_1,P_1)+Y_2(Z_2,P_2)&\to& V(e_1,r_1)+V(e_2,r_2)+Y_1(Z_1,P_1')+Y_2(Z_2,P_2').
\ea
Here, $P_i, P'_i$ are the four-momenta of the incoming and scattered ions, and
$e,e_i$ and $r,r_i$ are the polarization four-vectors and four-momenta and of
the created vector mesons, which obey the transversality conditions
$e(r)r=e_i(r_i)r_i=0$.
To describe the peripheral kinematics, it is convenient to introduce the
light-cone four-momenta $p_i$ as linear combinations of the incoming-ion
four-momenta $P_i$:
\ba
p_1=P_1-\lambda P_2,\qquad p_2=P_2-\eta P_1,\qquad  p_1^2=p_2^2=0, \qquad
P_1^2=m_1^2,\qquad P_2^2=m_2^2,\qquad 2P_1P_2\approx 2p_1p_2=s\gg m_i^2,
\ea
where $m_i$ are the masses of the ions.
In the calculation of the differential cross section, the effects of the
off-mass-shell-ness of the exchanged photons and vector meson must be taken
into account.
Our approach is based on taking the constituent quarks to be QED fermions.
The additional factors for the QED amplitudes associated with the color and
charge of the quarks will be discussed later. 

The details of the wave function of the bound state were discussed in
Ref.~\cite{GPS}.
For our approach, only one structure $R$ is relevant.
The virtual-photon polarization effects in the process
$\gamma^\star gg \to \psi$ were considered in Ref.~\cite{KNZ}.
The analysis of the inclusive annihilation of heavy quarkonium beyond the
Born approximation of QCD was presented in Ref.~\cite{BBL}.
In our work, we shall obtain the differential cross sections for the creation
of one or two vector mesons considered as bound states of the relevant quarks.


\section{Matrix elements of the $2\to 3$ processes}

To lowest order in perturbation theory, there are two sets of Feynman
diagrams involving three virtual-photon exchanges (see Fig.~1).
The contribution of each of them to the total cross section has the form
$\sigma\sim \sigma_0(aL^2+bL+c)$, with $L=\ln(s/M^2)$ being the ``large"
logarithm.
The interference of the relevant amplitudes only contributes terms devoid of
the ``large" logarithm.
Below, we restrict ourselves to the consideration of just one of the
amplitudes, which corresponds to exchanges of one virtual photon with one ion
$Y_1(Z_1,P_1)$ and of two virtual photons with the other ion $Y_2(Z_2,P_2)$.
Using the prescriptions proposed in Ref.~\cite{ECHAJA} for evaluating the
matrix element of the peripheral process of single vector meson production, we
obtain
\ba
M^{Y_1Y_2\to Y_1Y_2V}=\frac{(4\pi\alpha Z_1)(4\pi\alpha Z_2)^2}{q_1^2}\biggl(\frac{2}{s}\biggr)^3sN_1
\int\frac{d^4q}{(2\pi)^4q^2q_3^2}s^2N_2sF C^V ,
\ea
where the factor $C^V$ accounts for the color and charge of the quarks and
$N_1$, $N_2$, and $F$ are given by the following expressions:
\ba
N_1&=& \frac{1}{s}\bar{u}(P_1')\hat{p_2}u(P_1), \nn\\
N_2&=&\frac{1}{2s^2}\bar{u}(P_2')\biggl[\hat{p}_1\frac{\hat{p}_2-\hat{q}+m_2}{(p_2-q)^2-m_2^2}\hat{p}_1+
\hat{p}_1\frac{\hat{p}_2-\hat{q}_3+m_2}{(p_2-q_3)^2-m_2^2}\hat{p}_1\biggr]u(P_2), \,\,\\
%
%
F&=&\frac{1}{s}\frac{M {\cal{A}}}{2}\frac{1}{4} \Tr O^{\mu\nu\lambda}(\hat{p}+M_1)\hat{e}p_{1\mu}p_{2\nu}p_{2\lambda}.
\ea
%
Here, $M$ is the mass of the created vector meson, and $q_3 = q_2-q$.
The coupling constant ${\cal{A}}$, which we shall specify below, measures the
strength of the interaction of the vector meson with incoming photons.
The quantity $\bar{u}(q) O^{\mu\nu\lambda}v(q_+)$ is the matrix element of the
subprocess $3\gamma\to q\bar{q}$ depicted in Fig.~1.
So, we have
\ba
O^{\mu\nu\lambda}p_{1\mu}p_{2\nu}p_{2\lambda}&=&\hat{p}_1 \frac{\hat{q}_- -\hat{q}_1+m}{D_1}\biggl[\frac{1}{D_3}\hat{p}_2 (-\hat{q}_+ +\hat{q}_3+m)\hat{p}_2 +
\frac{1}{D_2}\hat{p}_2(-\hat{q}_+ +\hat{q}_2+m)\hat{p}_2\biggr] \nn \\
&&{}+\biggl[\frac{1}{D_2}\hat{p}_2(\hat{q}_- -\hat{q}+m)\hat{p}_2+\frac{1}{D_3}\hat{p}_2(\hat{q}_- -\hat{q}_3+m)\hat{p}_2\biggr]
\frac{-\hat{q}_+ +\hat{q}_1+m}{D_1}\hat{p}_1 \nn \\
&&{}+\frac{1}{D_2 D_3}[\hat{p}_2(\hat{q}_--\hat{q}+m)\hat{p}_1(-\hat{q}_++\hat{q}_3+m)\hat{p}_2+
\hat{p}_2(\hat{q}_--\hat{q}_3+m)\hat{p}_1(-\hat{q}_++\hat{q}_2+m)\hat{p}_2],
\ea
where $q_\pm$ are the four-momenta of the created quarks and the
denominators are given by the expressions
$D_1=(q_--q_1)^2-m^2+i0$, $D_{2}=(-q_-+q)^2-m^2$, and
$D_{3}=(-q_++q_{3})^2-m^2$.
Following the rules for the construction of the matrix element of the
quark-antiquark bound state \cite{Adkins}, we must put $q_+=q_-=p/2$ in this
expression.
Let us now introduce the Sudakov parametrization of the loop momenta and the
momenta of the quarks and virtual photons:
\ba
q_1&=&\beta_1 p_1+q_{1\bot},\qquad
q_{3}=\alpha_{3}p_2+\beta_{3}p_1+q_{3\bot},\qquad
q=\alpha p_2+\beta p_1+q_{\bot}, \nn \\
q_{\pm}&=&\alpha_{\pm}p_2+\beta_{\pm}p_1+q_{\pm\bot},\qquad
d^4q=\frac{s}{2}d\alpha d\beta d^2\vec{q}.
\label{sudakovParametr}
\ea
From four-momentum conservation and the on-mass-shell conditions of the quarks,
we have
\ba
\beta_+&=&\beta_-=\frac{1}{2}\beta_1,\qquad
\alpha+\alpha_3=\alpha_2; \nn\\
\vec{q}_1+\vec{q}_2 &=& \vec{q}_1+\vec{q}+\vec{q}_3=\vec{p},\qquad
s\alpha_{\pm}=\frac{1}{2\beta_1}(\vec{p}^2+4m^2).
\ea
The expressions for the denominators can now be rewritten as:
\ba
D_1&=&-\vec{q}_1^2+\vec{p}\vec{q}_1-\frac{1}{2}(\vec{p}^2+4m^2)
= - \frac{1}{2}(\vec{q}_1^2+\vec{q}_2^2+M^2) = -\frac{1}{2}R, \nn \\
D_2&=&-\vec{q}^2+\vec{p}\vec{q}-\frac{1}{2}s\beta_1\alpha + i0, \nn \\
D_3&=&-\vec{q}_3^2+\vec{p}\vec{q}_3-\frac{1}{2}s(\alpha-\alpha_2)\beta_1 + i 0.
\ea

Performing the integration over the component $\beta_2$ of the loop momentum,
we obtain for $N_2$
\ba
s\int d \beta N_2=\frac{1}{s}\bar{u}(P_2')\hat{p}_1u(P_2)
                \int\limits_{-\infty}^\infty s d \beta \biggl[\frac{1}{s\beta+a+i0}+\frac{1}{-s\beta+b+i0}\biggr]=
(-2\pi i)\frac{1}{s}\bar{u}(P_2')\hat u(P_2)=-2\pi i N_2'.
\ea
Summing the squared moduli of $N_1$ and $N_2'$ over the spin states,
we obtain
\ba
\sum|N_1|^2=2,
\qquad \sum|N_2'|^2  = 2.
\ea
In a similar way, the integral over the Sudakov variable $\alpha$ gives us
\ba
\int\limits_{-\infty}^\infty s d \alpha\biggl(\frac{1}{D_2}+\frac{1}{D_3}\biggr)=(-2\pi i)\frac{2}{\beta_1}.
\ea
As a result for the contribution of these terms to the matrix element, we have
\ba
\frac{8(4\pi\alpha Z_1)(4\pi\alpha Z_2)^2}{\vec{q}_1^2}N_1N_2'\frac{(2\pi i)^2}{4(2\pi)^4}\frac{1}{D_1}\frac{M_1{\cal{A}}}{2}
\int\frac{d^2\vec{q}}{\pi\vec{q}^2\vec{q}_3^2}
\frac{1}{4} \Tr Q ,
\ea
with the trace
\ba
\frac{1}{4} \Tr Q&=&
\frac{1}{4} \Tr[\hat{p}_2(-\frac{1}{2}\hat{p}+\hat{q}_1+m)\hat{p}_1
               -\hat{p}_1(\frac{1}{2}\hat{p}-\hat{q}_1+m)\hat{p}_2](\hat{p}+M)\hat{e} \nn\\
							 &=& \frac{1}{4} \Tr[\hat{p}_2 \hat{q}_1 \hat{p}_1 + \hat{p}_1 \hat{q}_1 \hat{p}_2] M \hat{e} =  sM (\vec{q}\vec{e}).
\ea
The contribution of the last two terms may be found using the relation
\ba
\int\limits_{-\infty}^\infty s d \alpha\frac{1}{D_2D_3}=(-2\pi i)\frac{2}{\beta_1}\frac{1}{D},
\ea
where $D = R-4\vec{q}\vec{q}_3$.
The calculation of the relevant trace leads to
\ba
&&\frac{1}{4}\Tr \hat{p}_2[(\hat{q}_{-\bot}-\hat{q}_{2\bot}+m)(-\hat{q}_{+\bot}+\hat{q}_{3\bot}-m)
           +(\hat{q}_{-\bot}-\hat{q}_{3\bot}+m)(-\hat{q}_{+\bot}-\hat{q}_{2\bot}-m)]
(\hat{p}+M)\hat{e} \nn \\
&&{}= -2m \frac{1}{4}\Tr \hat{p}_2 \hat{q}_1 \hat{p}\hat{e}=-s \frac{M}{2}\beta_1(\vec{q}_1\vec{e}).
\ea
Here, the expression contained within the square brackets is
$2m(\hat {q}_3 +\hat{q}_2 -\hat {q}_{+} -\hat {q}_{-}) = -2m \hat{q}_1$, the
replacement $(\hat {p} +M) \to \hat {p}$ is due to the relation $p e =0$, and
$p_2 p = \frac{1}{2}s\beta_1$.
Furthermore, we have
\ba
\frac{1}{D}-\frac{1}{R} =\frac{4 \vec{q}(\vec{q}_2-\vec{q})}{R D}.
\ea
%
We thus obtain
\ba
\int d\alpha F = \frac{(2\pi i) 8\vec{q}\vec{q}_3(\vec{e}\vec{q}_1)}{RD}\frac{M{\cal{A}}}{2}.
\ea

It is worth noting that the right-hand side of this equation vanishes in the
the limit where the transverse momenta of each of the three virtual photons
vanish, which is a consequence of gauge invariance.

The final result for matrix element corresponding to the Feynman diagram in
Fig.~1a is
\ba
M^{Y_1Y_2\to Y_1Y_2V}=M^2s\frac{2^7\pi^2Z_1Z_2^2\alpha^3}{\vec{q}_1^2} {\cal{A}} N_1N_2' (\vec{q}_1\vec{e})
\frac{1}{R}J(\vec{q}^2_2,R) C^V , 
\label{18}
\ea
%
with
\ba
J(\vec{q}_2^2,R) = \int\frac{d^2\vec{q}}{\pi\vec{q}^2\vec{q}_3^2}
    \frac{(\vec{q}\vec{q}_3)}{R-4(\vec{q}\vec{q}_3)},
\ea
where $R=\vec{q}_1^2 +\vec{q}_2^2 +M^2$ and $\vec{q}_3 =\vec{q}_2 - \vec{q}$.

As shown in Appendix~A, the function $J$ has the form
\ba
J(\vec{q}^2_2,R) = \frac{1}{2\vec{q_2}^2-R} \left[2\ln\frac{R}{2\vec{q}_2^2} - \ln\frac{R-\vec{q_2}^2}{\vec{q_2}^2} \right]
= \frac{1}{\vec{q}_1^2+M^2} {\cal{L}}(x_1),
\ea
where
\ba
{\cal{L}}(x)= \frac{1}{x-1}\ln\!\frac{(1+x)^2}{4x},\qquad
 x_{1,2}=\frac{\vec{q}^2_{2,1}}{\vec{q}_{1,2}^2+M^2}.
\ea

\section{Differential cross section of single vector meson production}

The Feynman diagrams contributing to the process of single vector meson
production are shown in Fig.~1.
The phase space volume has the form
\ba
d\Gamma_3=\frac{(2\pi)^4}{(2\pi)^9}\frac{d^3P_1'}{2E_1'}\frac{d^3P_1'}{2E_1'}\frac{d^3P_1'}{2E_1'}\delta^4(P_1+P_2-P_1'-P_2'-p).
\ea
We introduce the factor
$d^4q_1\delta^4(p_1-q_1-p_1') \, d^4q_2\delta^4(p_2-q_2-p_2')$
in terms of Sudakov variables and use the standard Sudakov parametrization
of Eq.~(\ref{sudakovParametr}) \cite{ECHAJA}.
Allowing for the vector meson to be unstable, the three-particle phase space
volume then becomes
\ba
d\Gamma_3=\frac{1}{4s}\frac{(2\pi)^4}{(2\pi)^9}\pi^2\frac{d^2\vec{q}_1d^2\vec{q}_2}{\pi^2}\frac{d\beta_1}{\beta_1}d p^2 R(p^2),
\ea
where $p^2=s\alpha_2\beta_1-(\vec{q}_1+\vec{q}_2)^2$ and we replace the delta
function by a Breit-Wigner resonance,
\ba
\delta(p^2-M^2)\to R(p^2)
=\frac{1}{\pi}\frac{\Gamma M}{(p^2-M^2)^2+M^2\Gamma^2},
\ea
where $M\approx 2 m_q$ and $\Gamma$ are the mass and the total decay width of
the vector meson resonance and $m_q$ is the mass of the bound quarks.
The quantity $p^2=(P_1+P_2-P_1'-P_2')^2$ may be associated with the missing mass
in the process of single vector meson production.

With the matrix element of Eq.~(\ref{18}), we now have
\ba
M^{Y_1Y_2\to Y_1Y_2V}&=&s\frac{2^7\pi^2 M^2 \alpha^3 Z_1 Z_2  N_1N_2}{R}{\cal{A}} Z,
\ea
where
\ba
Z=   \left[ \frac{Z_2(\vec{q_1}\vec{e})}{\vec{q}_1^2(\vec{q}_1^2+M^2)} {\cal {L}}(x_1)
+	 \frac{Z_1(\vec{q_2}\vec{e})}{\vec{q}_2^2(\vec{q}_2^2+M^2)} {\cal {L}}(x_2) \right] C^V.
\ea
The relevant differential cross section is
\ba
\frac{d\sigma^{(1)}}{dp^2} =\frac{ 2^6 \pi (Z_1 Z_2)^2 \alpha^6 {\cal{A}}^2 M^4 }{R^2} Z^2
 \frac{ d^2q_1 d^2q_2^2}{\pi^2}
 \frac{d\beta_1}{\beta_1}  R(p^2).
\ea
The interference term in $Z^2$ is canceled by the average over the azimuthal
angle.
For the case of extremely small transverse momentum $|\vec{q}_1|$ of ion $Y_1$,
a modification of this formula is necessary, which consists in replacing
\ba
\frac{1}{(\vec{q}_1^2)^2}\frac{d \beta_1}{\beta_1} \to \frac{1-\beta_1}{(\vec{q}_1^2+m_1^2\beta_1^2)^2}\frac{d \beta_1}{\beta_1},
\ea
and a similar replacement for small values of $|\vec{q}_2|$.
This leads to the so-called Weizs\"acker-Wiliams enhancement factor in the
total cross section:
\ba
\frac{d\sigma^{(1)}}{dp^2} = \sigma_0 R(p^2)[Z_2^2(L_1^2-5L_1)+Z_1^2(L_2^2-5L_2)+c(Z_1^2 +Z_2^2)+O(\frac{m_1^2}{s}, \frac{m_2^2}{s})],
\ea
where
\ba
\sigma_0&=&\frac{32 \pi(Z_1Z_2\alpha^3)^2 A^2}{M^2}(1-\ln2),
\nn\\
L_{1,2}&=&\ln\frac{s}{m_{1,2}^2},
\nn \\
c&=&2\int\limits_0^1\frac{dx}{x}\biggl[-\frac{5+x}{2(1-x)^3}+\frac{5}{2}
+\biggl(\frac{1+2x}{(1-x)^4}-1\biggr)\ln\frac{1}{x}\biggr]\approx 10.4565.
\ea
%

\section{Differential cross section of two vector meson production}

In the case of two vector mesons in the final state, we must consider the two
mechanisms due to the sets of Feynman diagrams indicated Figs.~2a and 2b.

The phase space volume of the four-particle final state is
\ba
d\Gamma_4=\frac{(2\pi)^4}{(2\pi)^{12}}\pi^3\frac{d\beta_1}{\beta_1}\frac{d\beta'}{\beta'}
\frac{d^2\vec{q}_1d^2\vec{q}'d^2\vec{q}_2}{\pi^3}
d r_1^2 d r_2^2 R(r_1^2)R(r_2^2)\frac{1}{8s},
\ea
where
$r_1^2=s\alpha'\beta_1-(\vec{q}_1-\vec{q}')^2$ and
$r_2^2=s\alpha_2\beta'-(\vec{q}_2+\vec{q}')^2$.
Also here, we assume that $m_{1,2}^2/s\ll \beta'\ll \beta_1\ll 1$.
Let us introduce another auxiliary four-vector:
$q'=\alpha'p_2+\beta'p_1+q'_\bot$.

The matrix element of the process of two vector meson production mediated by
a virtual vector meson (see Fig.~2a) has the form
\ba
M_a^{Y_1 Y_2\to Y_1 Y_2 V_1 V_2}=s
 \frac{(Z_1Z_2 \alpha^2)^2  N_1 N_2  {\cal{A}}_1 {\cal{A}}_2  (M_{V_1} M_{V_2} )^2 g^2 \pi^2 2^9 }{\vec{q}^2+M_V^2}
T (\vec{q}\, \vec{e}_1)(\vec{q} \,\vec{e}_2) C_2^{V_1} C_2^{V_2},
\ea
where $T$ is given by the expression
\ba
T = \frac{ J(q_1^2,R_1) J(q_2^2,R_2)}{R_1R_2} + \frac{ J(q_1^2,\bar{R}_1) J(q_2^2,\bar{R}_2)}{\bar{R}_1\bar{R}_2},
\ea
with $R_1 = q^2+q_1^2+M_1^2$,
$R_2 = q^2+q_2^2+M_2^2$,
$\bar{R}_1 = q^2+q_1^2+M_2^2$, and
$\bar{R}_2 = q^2+q_2^2+M_2^2$.
Here, $M_V$ is the mass of the virtual vector meson, $g$ is its coupling
constant, 
and $M_{1,2}$ are the masses of the created vector mesons.

The cross section of the process of two vector meson production is
\ba
\frac{d\sigma^{(2)}_a}{d r_1^2 d r_2^2}=s
\frac{  2^6 \alpha^8  (M_1 M_2 Z_1 Z_2)^4 }{(\vec{q}^2+M_V^2)^2}
g^4 ({\cal{A}}_1 {\cal{A}}_2)^2 R(r_1^2) R(r_2^2) |T|^2
  (\vec{q}\vec{e}_1)^2(\vec{q}\vec{e}_2)^2 ( C_2^{V_1} C_2^{V_2})^2
\frac{d^2\vec{q}_1 d^2\vec{q}_2 d^2\vec{q}}{\pi^3}
\frac{d\beta_1}{\beta_1}\frac{d\beta}{\beta}.
\ea
After integration over $d^2\vec{q}$, we obtain:
\ba
\frac{d\sigma^{(2)}_a}{d r_1^2 d r_2^2}=\frac{8 Z_1^4Z_2^4 g^4 \alpha^8 }{\pi} (1+2 \cos^2\theta_{12})
\frac{d\beta_1}{\beta_1}\frac{d\beta}{\beta}
 ({\cal{A}}_1 {\cal{A}}_2)^2  \left( \frac{ M_{V_1}^2 M_{V_2}^2 }{ M_V^5} \right)^2
  R(r_1^2) R(r_2^2) ( C_2^{V_1} C_2^{V_2})^2 \frac{d\vec{q}_1^2 d\vec{q}_2^2}{M_V^4}P(\frac{\vec{q}_1^2}{M_V^2},\frac{\vec{q}_2^2}{M_V^2}),
\label{eq:P}
\ea
where $\cos \theta_{12}=\cos \widehat{ \vec{e}_1 \vec{e}_2 } $ and
\ba
P(\frac{\vec{q}_1^2}{M_V^2},\frac{\vec{q}_2^2}{M_V^2})=\int\limits_0^{\infty}
\frac{d\vec{q}^2}{(\vec{q}^2+M_V^2)^2}\frac{(\vec{q}^2)^2}{M_V^2}|T M_V^8|^2
\ea
is evaluated in Appendix B yielding the numerical values presented in Table~I.

For the case of two-gluon exchange (see Fig.~2b), the matrix element has the
form
\ba
M_b^{Y_1 Y_2\to Y_1 Y_2 V_1 V_2}=i s
\frac{ 2^{12} \pi^3  (Z_1Z_2 \alpha^2\alpha_s^2) ( M_{V_1} M_{V_2})^2 N_1 N_2}{ \vec{q}_1^2 \vec{q}_2^2 }{
\cal{A}}_1 {\cal{A}}_2 Q  C_1^{V_1} C_1^{V_2},
\ea
where
\ba
Q = \frac{I(R_1,R_2)}{R_1R_2}(\vec{e}_1\vec{q}_1)(\vec{e}_2\vec{q}_2)  +
\frac{I(\bar{R}_1,\bar{R}_2)}{\bar{R}_1\bar{R}_2}(\vec{e}_1\vec{q}_2)(\vec{e}_2\vec{q}_1).
\label{eq:I}
\ea
The quantity $I(R_1,R_2)$ is given in the Appendix~A.
For the cross section, we may write
\ba
\frac{d\sigma^{(2)}_b}{d r_1^2 d r_2^2}=
\frac{2^{12} \pi (Z_1Z_2 \alpha^2\alpha_s^2)^2  ( M_{V_1} M_{V_2})^4 }{(\vec{q}_1^2 \vec{q}_2^2 )^2}
R(r_1^2)R(r_2^2) Q^2
\frac{d\beta_1}{\beta_1}
\frac{d\beta}{\beta}({\cal{A}}_1 {\cal{A}}_2 )^2
\frac{d^2\vec{q}_1d^2\vec{q}_2d^2\vec{q}}{\pi^3} ( C_2^{V_1} C_2^{V_2})^2.
\ea
Performing the integration over $d^2\vec{q}$, we obtain
\ba
\frac{d\sigma^{(2)}_b}{d r_1^2 d r_2^2}= \frac{2^{10}\pi(z_1z_2\alpha^2\alpha_s^2)^2 dx_1 dx_2}{x_1 x_2 M_{V_1}M_{V_2}}
R(r_1^2)R(r_2^2)\frac{d\beta_1}{\beta_1}\frac{d\beta}{\beta}({\cal{A}}_1 {\cal{A}}_2 )^2 ( C_2^{V_1} C_2^{V_2})^2
\biggl[\Phi +\bar {\Phi}+2\cos^2\theta_{12} G \biggr]W,  
\label{eq:Phi}
\ea
where the functions $\Phi$, $\bar{\Phi}$, and $G$ are evaluated in Appendix~B
yielding the numerical values presented in Table~II,
\ba
W=\frac{d\vec{q}_1^2}{\vec{q}_1^2}\frac{d\vec{q}_2^2}{\vec{q}_2^2}
\ea
is the Weizs\"acker-Williams enhancement factor, and
\ba
x_{1,2} = \frac{\vec{q}_{1,2}^2}{M_{V_1}M_{V_2}}.
\ea
In the region of small values of $\vec{q}^2_{1,2}$, we must replace
$\vec{q}_1^2 \to\vec{q}^2_1+\beta_1^2m_1^2$,
$\vec{q}_2^2 \to\vec{q}^2_2+\alpha_2^2m_1^2$.

We note that the interference term of the two amplitudes is absent, so that
$|M_b^{(2)} +M_a^{(2)}|^2 = |M_b^{(2)}|^2 + |M_a^{(2)}|^2$.

\section{Conclusion}

We studied the production of one or two vector mesons in peripheral heavy-ion
collisions at high energies.
In the case of $Z\alpha > \alpha_s$, a simplified version of a general theory
\cite{BBL} can be used to lowest order of QED and QCD that is based on the
subprocesses $\gamma^\star\gamma^\star g\to V$ and $\gamma^\star gg \to V$.
In our study of two vector meson production, a vector meson can also appear
virtually as an intermediate state.
In this case, it is important to replace it by the relevant vector reggeon
state, with the Regge trajectory
\ba
\alpha_V((\vec{q}')^2)=\alpha_V(0)+\alpha'_V(0)(\vec{q}')^2\approx \alpha_V(0)\approx 1/2.
\ea
This results in an additional factor
\ba
\biggl(\frac{p^2}{s_0}\biggr)^{2(\alpha_V(0)-1)}\sim \frac{s_0}{p^2},
\ea
where $s_0\sim 1$~GeV and $p^2$ is the missing mass square, in the cross
section $d\sigma^{(2)}_a$.

When constructing the invariant mass square of the decay products of one vector
meson, $p^2=(P_1+P_2-P_1'-P_2')^2=s\beta_1\alpha_2-(\vec{q}_1+\vec{q}_2)^2$,
we take into account that the part $s\beta_1\alpha_2$ is the combination
$(\sum E_i)^2-(\sum p_{iz})^2$, with $E_i$ and $p_{iz}$ being the energies and
$z$ components of the three-momenta of the decay products, while the part
$-(\vec{q}_1+\vec{q}_2)^2$ is the contribution from the transverse components
$-(\sum p_{i\bot})^2$.
Here, it is understood that the $z$ direction is taken along the beam axis in
the center-of-mass frame.
Such is the case for two jet production with
$r_1^2=(P_1-P_1'-q')^2=s\alpha'\beta_1-(\vec{q}_1-\vec{q}')^2$ and
$r_2^2=(P_2-P_2'+q')^2=s\alpha_2\beta'-(\vec{q}_2+\vec{q}')^2$.

The coupling constant ${\cal{A}}$ of the meson-photon interaction in the case
of single vector meson production, appearing in Eq.~(5), is given by
${\cal{A}} = \alpha^{3/2}/(2\sqrt{\pi})$ for ortho-positronium and by
${\cal{A}}_i = 2 f_{V_i}/ M_{V_i}$, with $f_\rho = f_\omega = 0.21$~GeV and
$f_\psi = 0.38$~GeV, for the $\omega$, $\rho$, and $J/\psi$ mesons,
respectively (see Ref.~\cite{klevanski} for details).

In the case of single vector meson production, the color and charge factors
are $C^V  = 3\sum\limits_{u,\,d} Q^3_q = \frac{7}{3}$ for the $\rho$ and
$\omega$ mesons, and $C^V  = \frac{8}{3}$ for the $J/\psi$ meson.
In the case of two vector meson production through the mechanism shown in
Fig.~2a, we have $C_2^{V_1} = 3\sum Q_q^2 = \frac{5}{3}$ for the $\rho$ and
$\omega$ mesons, and $C_2^{V_1}  = \frac{4}{3}$ for the $J/\psi$ meson.
In the case of the mechanism shown in Fig.~2b, we have
$C_1^{V_1} = 3\sum Q_q = 1$ for the $\rho$ and $\omega$ mesons, and
$C_1^{V_1} = 3\frac{2}{3} = 2$ for the $J/\psi$ meson.

Note that the mechanism involving single $\gamma^*$ exchange (see Fig.~2b)
yields a $L^4$ enhancement, whereas double $\gamma^*$ exchange (see Fig.~2a)
only produces a $L^2$ enhancement.
For $pp$ collisions at the LHC, the ``large" logarithm is as large as
$L = \ln\frac{s}{m^2}\approx7$.

We do not consider gluon exchange between heavy ions and vector mesons to avoid
channels with ion exitation.

\appendix

\section{}
\label{AppendixA}

In this section, we shall explain how to calculate the two-dimensional
Euclidean integrals appearing in Eqs.~(\ref{18}) and (\ref{eq:I}),
\ba
J (\vec{q}_2^2,R) = \int\frac{d^2\vec{k}}{\pi}\frac{\vec{k}(\vec{q}_2-\vec{k})}{\vec{k}^2(\vec{q}_2-\vec{k})^2 D},\qquad
I(R_1, R_2) = 4 \int\frac{d^2\vec{k}}{\pi}\frac{(\vec{k}(\vec{q}-\vec{k}))^2}{\vec{k}^2(\vec{q}-\vec{k})^2 D_1 D_2},
\ea
where $D = R-4 \vec{k}(\vec{q}-\vec{k})$, $R = \vec{q}_1^2+\vec{q}_2^2 +M^2$, $D_i = R_i - 4 \vec{k}(\vec{q}-\vec{k})$, and $R_i = q_i^2+q^2+M_i^2$.
For the first one, we have
\ba
4 J = \int \frac{d^2 \vec{k}}{\pi}\frac{R-D}{\vec{k}^2(\vec{q_2}-\vec{k})^2 D} =
  \lim_{\lambda\to 0}\biggl[ R J_1 - J_0 \biggr],
\ea
with
\ba
J_1 = 	\int \frac{d^2 \vec{k}}{\pi}\frac{1}{((\vec{q_2}-\vec{k})^2+\lambda^2)(\vec{k}+\lambda^2) D}, \qquad
J_0 = \int \frac{d^2 \vec{k}}{\pi}\frac{1}{(k^2+\lambda^2)((\vec{q_2}-\vec{k})^2+\lambda^2)}.
\ea
Using Feynman's trick of joining the denominators,
\ba
\frac{1}{ab} = \int\limits_0^1 dx\frac{1}{(x a +\bar{x}b)^2},
\ea
with $\bar{x}=1-x$,
we obtain for $J_0$:
\ba
J_0 = \int\limits_0^1 d x \int
 \frac{d^2 \vec{k}}{\pi}\frac{1}{[(\vec{k}-\vec{q_2}^2 x)^2+\vec{q_2}^2x\bar{x}+\lambda^2]^2} =
\int\limits_0^1\frac{dx}{\vec{q}^2_2 x\bar{x}+\lambda^2} = \frac{2}{\vec{q}_2^2}\ln\frac{\vec{q}_2^2}{\lambda^2}.
\ea
For $J_1$, we have
\ba
4 J_1=\int\limits_0^1d x\int\limits_0^1
  \frac{y d y}{[\vec{q_2}^2 x\bar{x}y^2 + \frac{\bar{y}}{4}(R-\bar{y}\vec{q}_2^2)+y\lambda^2]^{2}},
\ea
where $\bar{y} = 1-y$.
Introducing the variable $t=1-2x$, we may cast this into the form:
\ba
J_1=4\int\limits_0^1 y d y\int\limits_0^1  \frac{d t}{(A-B t^2 )^2},
\ea
where
\ba
A = \vec{q}_2^2 y^2 + \bar{y}(R-\vec{q_2}^2\bar{y})+4 y \lambda^2, \qquad
 B = \vec{q}_2^2 y^2.
\ea
Performing the integration over $t$, we obtain
\ba
J_1 = \int\limits_0^1\frac{ y d y}{A^{3/2}B^{1/2}}\Bigl[\frac{2(AB)^{1/2}}{A-B} + \ln\frac{A^{1/2}+B^{1/2}}{A^{1/2}-B^{1/2}}\Bigr]
 = (I_1+I_2)\Bigl(\frac{1}{\vec{q}^2}\Bigr)^2,
\label{eq:J1}
\ea
where
\ba
I_1 = 2  \int\limits_0^1\frac{ y d y}{[\bar{y}(\rho-\bar{y}) + 4 y \sigma]T},
\qquad
I_2 = \int\limits_0^1 \frac{d y}{T^{3/2}} \ln\frac{T^{1/2}+y}{T^{1/2}-y},
\ea
with
\ba
T = y^2+\bar{y}(\rho-\bar{y}),\qquad
\rho = \frac{R}{\vec{q}^2},\qquad
\sigma = \frac{\lambda^2}{\vec{q}^2}.
\ea

The first integral $I_1$ contains an infrared singularity.
Introducing the small parameters $\sigma$ and $\epsilon$, with
$\sigma\ll\epsilon\ll1$, we rewrite it as
\ba
I_1 &=& 2 \int\limits_0^{1-\epsilon}\frac{ y d y}{\bar{y}(\rho-\bar{y})[\rho-1+y(2-\rho)]}
+  2  \int\limits_{1-\epsilon}^1\frac{  d y}{\bar{y} \rho+4 \sigma} \nn\\
&=&\frac{2}{\rho}\ln\frac{\rho}{4\sigma} - \frac{2}{\rho(\rho-1)}[(\rho-1)\ln(\rho-1)+\ln\rho].
\ea

For the second integral $I_2$, which is infrared finite, the substitutions
$T=t^2$, $y=\frac{t^2-a}{b}$, $a=\rho-1$, and $a+b=1$ yield
\ba
I_2&=& \frac{2}{b} \int\limits^1_{\sqrt{a}}\frac{dt}{t^2}
\ln\frac{tb+t^2-a}{tb-(t^2-a)}=\frac{2}{b} \int\limits^1_{\sqrt{a}}\frac{dt(1-t)}{t}
\Bigl[\frac{b+2t}{tb+t^2-a}-\frac{b-2t}{tb-t^2+a}\Bigr]
\nn\\
&=&4\int\limits^1_{\sqrt{a}} dt \frac{t^2+a}{t(1+t)(t^2-a^2)}
= \frac{2}{b} [\ln  a- 2 \ln 2 + \frac{a+1}{a}\ln(a+1)].
\ea

The total answer for $J_1$ is
\ba
J_1 (\rho)= 2 \left(\frac{1}{\vec{q}^2}\right)^2\left[\frac{1}{\rho}\ln\frac{1}{\sigma}
+ \frac{2}{\rho(2-\rho)}( 2 \ln\rho - 2 \ln2 - \ln(\rho-1))\right].
\ea
For the sum $J = R J_1 -J_0$, we obtain
\ba
J (R,\vec{q}^2_2)= \frac{1}{2\vec{q}_2^2-R} \ln \frac{R^2}{4(R-\vec{q}^2_2)\vec{q}^2_2}
=\frac{1}{\vec{q}_1^2+M^2}\frac{1}{x-1}\ln\frac{(x+1)^2}{4x},
\ea
where $x=\vec{q}_2^2/(\vec{q}_1^2+M^2)$.

For the integral $I$, we have
\ba
4 I = \int\frac{d^2\vec{k}}{\pi}\frac{1}{\vec{k}^2(\vec{q}-\vec{k})^2}\frac{(R_1-D_1)(R_2-D_2)}{D_1D_2} =
     J_0 + \frac{R_2^2}{R_1 - R_2} J_1(R_2) - \frac{R_1^2}{R_1 - R_2} J_1(R_1),
\ea
where $J_1$ given in Eq.~(\ref{eq:J1}) and
$R_{1,2} = \vec{q}_{1,2}^2+\vec{q}^2+M_{1,2}^2$.
The infrared singularity is canceled using
\ba
 I = \frac{1}{R_1-R_2}\Biggl\{\frac{R_2}{2\vec{q}^2-R_2} \ln\frac{R_2^2}{4\vec{q}^2(R_2-\vec{q}^2)}
 - \frac{R_1}{2\vec{q}^2-R_1}\ln\frac{R_1^2}{4\vec{q}^2(R_1-\vec{q}^2)} \Biggr\}.
\ea
For the case of $R_1 =R_2 =R = \vec{q}_1^2 +\vec{q}^2 +M^2$, we have
\ba
I(R, R) = -\frac{\partial}{\partial R}\int\frac{d^2k}{\pi}\frac{\vec{k}(\vec{q}-\vec{k}) (R-D)}{\vec{k}^2(\vec{q}-\vec{k})^2 D} = -4 \frac{\partial}{\partial R}R J(R, \vec{q}^2) =
4\biggl[\frac{1}{R-\vec{q}_1^2}-\frac{2\vec{q}_1^2}{2\vec{q}_1^2-R}\ln\frac{R^2}{4\vec{q}_1^2(R^2-\vec{q}_1^2)}\biggr].
\label{A14}
\ea
For the cross section of single vector meson production, integrated over
$\vec{q}_1^2$ and $\vec{q}_2^2$, we have
\ba
\frac{d\sigma^{(1)}}{dp^2 R(p^2)} &=&\frac{16\pi(Z_1Z_2\alpha^3)^2 A^2}{M^2}
\int\limits_0^{\infty}\frac{dx}{(x+1)^2(x-1)^2}\ln^2 \biggl(\frac{(x+1)^2}{4x}\biggr)\biggl\{Z_1^2 \int\limits_{\frac{m_1^2}{s}}^1\frac{d\beta_1}{\beta_1}(1-\beta_1)
\int\limits_0^{\infty}\frac{dt \cdot t}{(1+t)^3(t+\beta_1^2 \rho_1^2)}
\nn\\
&&{}+(\beta_1 \to \alpha_2, Z_1 \leftrightarrow Z_2,\rho_1 = \frac{m_1^2}{M^2} \to \rho_2 = \frac{m_2^2}{M^2})\biggr\}.
\ea
Using
\ba
\int\limits_0^{\infty}\frac{dx}{(x^2-1)^2}\ln^2(\frac{(x+1)^2}{4x})=2(1-\ln2),
\ea
we obtain the expression given in Eq.~(31).

\section{}
\label{AppendixB}

In this section, we shall explain how to evaluate the integrals appearing in
Eqs.~(\ref{eq:P}) and (\ref{eq:Phi}) relevant for the mechanisms based on vector meson
(see Fig.~2a) and two-gluon (see Fig.~2b) exchange, respectively.

In the first case, we have
\ba
P(x_1, x_2; \rho_1, \rho_2)
=\int\limits_0^{\infty}\frac{dx \cdot x^2}{(x+1)^2}\tau^2,
\label{BP}
\ea
with
\ba
\tau=\frac{i(x_1, r_1) i(x_2, r_2)}{r_1 r_2} +\frac{i(x_1,\bar{r}_1) i(x_2, \bar{r}_2)}{\bar{r}_1 \bar{r}_2},
\ea
where
\ba
i(x,r) = \frac{1}{2x-r}\ln\frac{r^2}{4x(r-x)},
\ea
and
$r_1 = x + x_1 +\rho_1$,
$r_2 = x+x_2 +\rho_2$,
$\bar{r}_1 =x+x_1+\rho_2$,
$\bar{r}_2 =x+x_2+\rho_1$,
$\rho_1=M_{V_1}^2/M_V^2$, and
$\rho_2=M_{V_2}^2/M_V^2$.
Numerical values of $P(x_1,x_2;1,1)$, appropriate for the important case
$\rho_1=\rho_2=1$, are listed in Table~\ref{TableP}.

On the other hand, we have
\ba
\Phi(x_1, x_2; \rho_1, \rho_2)
&=&\int\limits_0^{\infty}dx\biggl(\frac{j(r_1, r_2)}{r_1 r_2}\biggr)^2,
\nonumber\\
\bar {\Phi}(x_1, x_2; \rho_1, \rho_2)&=&
\int\limits_0^{\infty}dx\biggl(\frac{j(\bar{r}_1, \bar{r}_2)}{\bar{r}_1 \bar{r}_2}\biggr)^2,
\nonumber\\
G(x_1, x_2; \rho_1, \rho_2)&=&
 \int\limits_0^{\infty}dx \frac{j(r_1, r_2) j(\bar{r}_1, \bar{r}_2)}{r_1 r_2 \bar{r}_1, \bar{r}_2},
\label{BFG}
\ea
where
\ba
j(r_1, r_2) = \frac{1}{r_1 -r_2}\biggl[\frac{r_2}{2x-r_2}\ln\frac{r_2^2}{4x(r_2-x)} - \frac{r_1}{2x-r_1}\ln\frac{r_1^2}{4x(r_1-x)}\biggr].
\ea
In the case $r_1 = r_2$, we have (see Eq.~(\ref{A14}))
\ba
j(r_1, r_1) = 4- \frac{8 x}{(x-1)^2}\ln\biggl(\frac{(x+1)^2}{4x}\biggr), \,\,\,\,x=\frac{\vec{q}_1^2}{\vec{q}^2+M^2}.
\ea
For $\rho_1=\rho_2$, we have
$\Phi(x_1, x_2; \rho_1, \rho_1)=\bar{\Phi}(x_1, x_2; \rho_1, \rho_1)
=G(x_1, x_2; \rho_1, \rho_1)$.
Numerical values for the important choice $\rho_1=\rho_2=1$ are listed in
Table~\ref{TableF}.

\begin{table}
  \begin{tabular}{|c|c|c|c|c|c|c|c|c|}
  \hline
    $x_1$ $\backslash$ $x_2$ & $0.1$    & $0.5$    & $1$   & $1.5$   & $2$   & $3$  & $4$  & $5$  \\
  \hline
    $0.1$ &         $0.076$ & $0.0047$ & $0.00049$ & $0.000082$ & $0.00004$ & $0.000046$ & $0.000045$ & $0.000038$ \\
  \hline
    $0.5$ &         $0.0047$ & $0.00036$ & $0.000049$ & $0.00001$ & $3.66 \cdot 10^{-6}$ & $2.51 \cdot 10^{-6}$  & $2.48 \cdot 10^{-6}$  & $2.23 \cdot 10^{-6}$  \\
  \hline
    $1$ &          $0.00049$ &  $0.000049$ & $9.0049 \cdot 10^{-6}$ & $2.34 \cdot 10^{-6}$ & $7.85 \cdot 10^{-7}$  & $2.52 \cdot 10^{-7}$ & $2.136 \cdot 10^{-7}$  & $2.03 \cdot 10^{-7}$ \\
  \hline
    $1.5$ &          $0.000082$ &  $0.00001$ & $2.34 \cdot 10^{-6}$ & $7.39 \cdot 10^{-7}$  & $2.78 \cdot 10^{-7}$  & $6.46 \cdot 10^{-8}$  &  $3.357 \cdot 10 ^{-8}$  & $2.85 \cdot 10^{-8}$ \\
  \hline
    $2$ &          $0.00004$ &  $3.66 \cdot 10^{-6}$ & $7.85 \cdot 10^{-7}$  & $2.78 \cdot 10^{-7}$ & $1.29 \cdot 10^{-7}$ & $4.59 \cdot 10^{-8}$ & $2.52 \cdot 10^{-8}$  & $1.76 \cdot 10^{-8}$ \\
  \hline
    $3$ &          $0.000046$ & $2.51 \cdot 10^{-6}$ & $2.52 \cdot 10^{-7}$  & $6.46 \cdot 10^{-8}$ & $4.59 \cdot 10^{-8}$ & $4.02 \cdot 10^{-8}$  & $3.21 \cdot 10^{-8}$  & $2.45 \cdot 10^{-8}$ \\
  \hline
    $4$ &         $0.000045$ & $2.48 \cdot 10^{-6}$ & $2.14 \cdot 10^{-7}$  & $3.36 \cdot 10^{-8}$  & $2.52 \cdot 10^{-8}$  & $3.21 \cdot 10^{-8}$  & $2.91 \cdot 10^{-8}$  & $2.35 \cdot 10^{-8}$  \\
  \hline
    $5$ &         $0.000038$  & $2.23 \cdot 10^{-6}$  & $2.029 \cdot 10^{-7}$  & $2.85 \cdot 10^{-8}$  & $1.76 \cdot 10^{-8}$  & $2.45 \cdot 10^{-8}$  & $2.35 \cdot 10^{-8}$  & $1.96 \cdot 10^{-8}$  \\
  \hline
  \end{tabular}
  \caption{Values of the function $P(x_1, x_2; 1,1)$, defined in
Eq.~(\ref{BP}), for different values of $x_1$ and $x_2$.}
  \label{TableP}
\end{table}
\begin{table}
  \begin{tabular}{|c|c|c|c|c|c|c|c|c|}
  \hline
    $x_1$ $\backslash$ $x_2$ & $0.1$    & $0.5$    & $1$   & $1.5$   & $2$   & $3$  & $4$  & $5$  \\
  \hline
    $0.1$ &         $1.67$ & $1.006$ & $0.623$ & $0.428$ & $0.314$ & $0.191$ & $0.129$ & $0.094$ \\
  \hline
    $0.5$ &         $1.006$ & $0.6142$ & $0.386$ & $0.268$ & $0.198$ & $0.123$  & $0.08399$  & $0.0615$  \\
  \hline
    $1$ &          $0.623$ &  $0.386$ & $0.246$ & $0.173$ & $0.129$  & $0.0808$ & $0.0559$  & $0.0414$ \\
  \hline
    $1.5$ &          $0.428$ &  $0.268$ & $0.173$ & $0.1221$  & $0.0918$  & $0.0583$  &  $0.04078$  & $0.0304$ \\
  \hline
    $2$ &          $0.314$ &  $0.198$ & $0.129$  & $0.0918$ & $0.0695$ & $0.0446$ & $0.0314$  & $0.0235$ \\
  \hline
    $3$ &          $0.191$ & $0.123$ & $0.0808$  & $0.0583$ & $0.0446$ & $0.029$  & $0.02073$  & $0.0157$ \\
  \hline
    $4$ &         $0.1297$ & $0.08399$ & $0.0559$  & $0.04078$  & $0.0314$  & $0.02073$  & $0.149$  & $0.01142$  \\
  \hline
    $5$ &         $0.094$  & $0.0615$  & $0.0414$  & $0.0304$  & $0.0235$  & $0.0157$  & $0.01142$  & $0.0088$  \\
  \hline
  \end{tabular}
  \caption{Values of the function
$\Phi(x_1, x_2; 1,1)=\bar\Phi(x_1, x_2; 1,1)=G(x_1, x_2; 1,1)$, defined in
Eq.~(\ref{BFG}), for different values of $x_1$ and $x_2$.}
  \label{TableF}
\end{table}
\acknowledgments
We are grateful to V.~Pozdnyakov for a discussion about the experimental
situation at the LHC.
We are grateful to \v {C}. ~Burd\'{\i}k  for the taking part in initial stage of work.
This work was supported in part by the German Federal Ministry for Education
and Research BMBF through Grant No.\ 05~HT6GUA and by the Helmholtz
Association HGF through Grant No.\ Ha~101.
The work of E.A.K. was supported in part by Russian Foundation for Basic
Research RFBR through Grant No.\ 01201164165 and the Heisenberg-Landau Grant
No.\ HLP-2012-11.
%


%
\clearpage
\begin{figure}
\centering
{\includegraphics[width=0.8\textwidth]{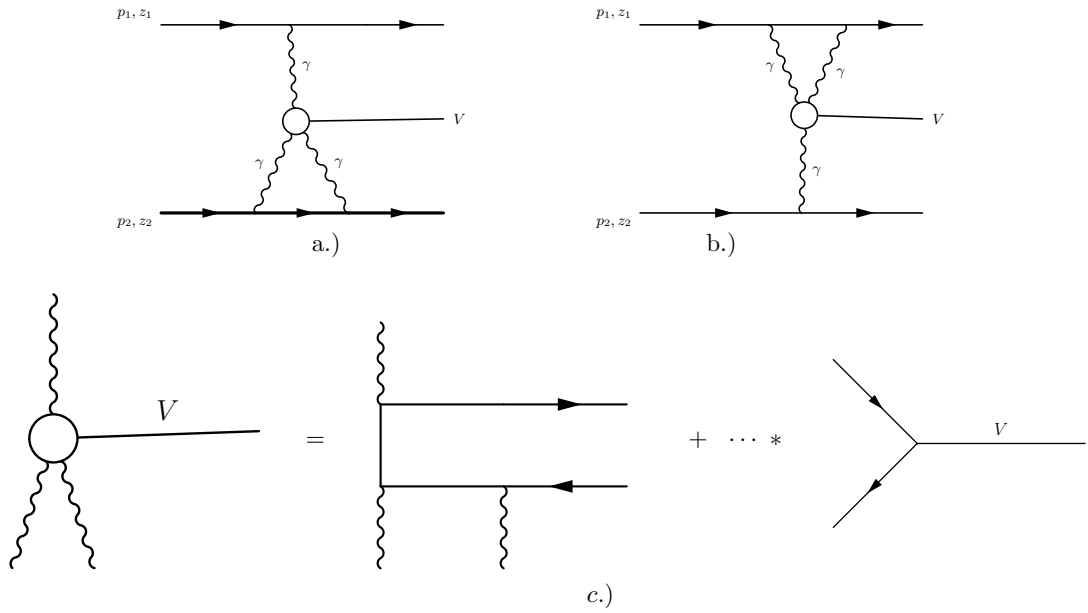}}
 \caption{\label{Fig:1ab}
Feynman diagrams pertinent to single vector meson production in peripheral
heavy-ion collisions.
}
\end{figure}
\begin{figure}
\centering
{\includegraphics[width=0.8\textwidth]{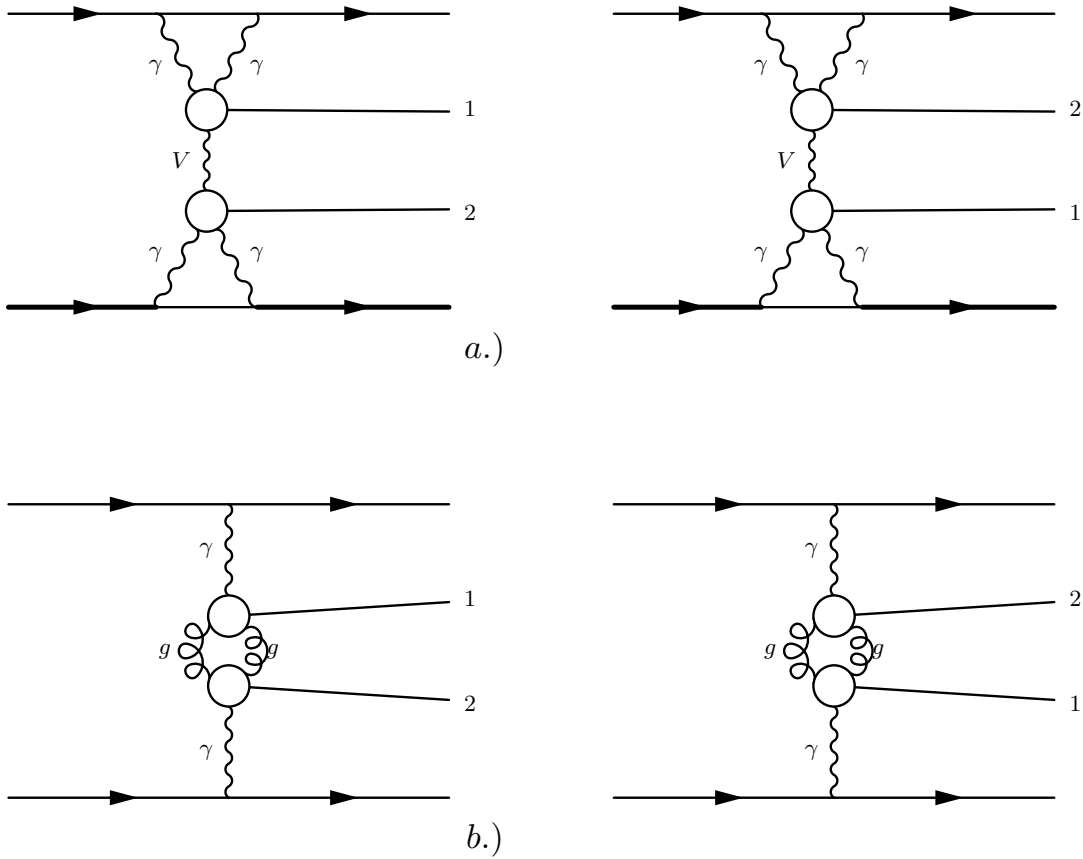}}
 \caption{\label{Fig:2ab}
Feynman diagrams pertinent to two vector meson production in peripheral
heavy-ion collisions via intermediate a) vector meson and b) two-gluon states.}
\end{figure}
%


\begin{thebibliography}{10}

\bibitem{ECHAJA}
  A.~B.~Arbuzov, V.~V.~Bytev, E.~A.~Kuraev, E.~Tomasi-Gustafsson, and
 Yu.~M.~Bystritskiy,
  Phys.\ Part.\ Nucl.\  {\bf 41}, 593 (2010).

\bibitem{GPS}
  I.~F.~Ginzburg, S.~L.~Panfil, and V.~G.~Serbo,
  Nucl.\ Phys.\ B {\bf 284}, 685 (1987);
  I.~F.~Ginzburg, S.~L.~Panfil, and V.~G.~Serbo,
  Nucl.\ Phys.\ B {\bf 296}, 569 (1988).

\bibitem{KNZ}
  E.~V.~Kuraev, N.~N.~Nikolaev, and B.~G.~Zakharov,
  JETP Lett.\  {\bf 68}, 696 (1998)
  [Pisma Zh.\ Eksp.\ Teor.\ Fiz.\  {\bf 68}, 667 (1998)]
  [hep-ph/9809539].

\bibitem{BBL}
  G.~T.~Bodwin, E.~Braaten, and G.~P.~Lepage,
  Phys.\ Rev.\ D {\bf 51}, 1125 (1995)
  [Erratum-ibid.\ D {\bf 55}, 5853 (1997)]
  [hep-ph/9407339].

\bibitem{Adkins}
  G.~S.~Adkins and Y.~Shiferaw,
  Phys.\ Rev.\ A {\bf 52}, 2442 (1995).

\bibitem{klevanski}
  M.~V.~Terentev,
  Sov.\ J.\ Nucl.\ Phys.\  {\bf 24}, 106 (1976)
  [Yad.\ Fiz.\  {\bf 24}, 207 (1976)];
  V.~B.~Berestetsky and M.~V.~Terentev,
  Sov.\ J.\ Nucl.\ Phys.\  {\bf 25}, 347 (1977)
  [Yad.\ Fiz.\  {\bf 25}, 653 (1977)].
  S.~P.~Klevansky,
  Rev.\ Mod.\ Phys.\  {\bf 64}, 649 (1992).

\end{thebibliography}
\end{document}